\documentclass[%
 aps,
 prl,
twocolumn,
superscriptaddress,
frontmatterverbose, 
showpacs,preprintnumbers,
nofootinbib,
 amsmath,
 amssymb,
floatfix,
latexsym,array,enumerate,letter,
]{revtex4-1}

\pdfoutput=1
 \usepackage{titlesec}
   \titleformat{\section}[runin]
   {\normalfont\bfseries}{\quad\thesection.}{0.5em}{} [.]
  \titlespacing*{\section}{0pt}{0.2\baselineskip}{\baselineskip}

\usepackage{slashed}
\usepackage[utf8]{inputenc}
\usepackage[english]{babel}    
\usepackage{tocvsec2}
\usepackage{amsthm} 
\pdfoutput=1
\allowdisplaybreaks
\usepackage{graphicx,color}
\usepackage[colorlinks=true,citecolor=blue,linkcolor=blue,urlcolor=blue]{hyperref}
\usepackage{comment}
\usepackage{appendix}

 \usepackage{graphicx, psfrag} 

\usepackage{footmisc}
\usepackage{bm}
\usepackage{dcolumn}
\usepackage{gensymb}
 \usepackage{float}
\restylefloat{table}
\usepackage{multirow}
 \usepackage{hyperref}
\hypersetup{
    colorlinks,
    citecolor=black,
    filecolor=black,
    linkcolor=blue,
    urlcolor=blue
}
\usepackage{epsfig}
\def\be{\begin{equation}}
\def\ee{\end{equation}}
\def\bea{\begin{eqnarray}}
\def\eea{\end{eqnarray}}
\def\ba{\begin{equation}\begin{aligned}}

\begin{document}
\title{Unified Interacting Quark Matter and its Astrophysical Implications}
\author{Chen Zhang}
\email{czhang@physics.utoronto.ca}
\affiliation{Department of Physics, University of Toronto, 60 St. George Street, Toronto, Ontario, M5S 1A7, Canada}
\affiliation{Department of Physics and Astronomy, University of Waterloo, Waterloo, Ontario, N2L 3G1, Canada}
\author{Robert B. Mann}
\email{rbmann@uwaterloo.ca}
\affiliation{Department of Physics and Astronomy, University of Waterloo, Waterloo, Ontario, N2L 3G1, Canada}


\begin{abstract}
We investigate interacting quark matter (IQM),  including the perturbative QCD correction and color superconductivity, for both up-down quark matter and strange quark matter. We first derive an equation of state (EOS)  unifying all cases by a simple reparametrization and rescaling, through which we manage to maximally reduce the number of degrees of freedom. We find, in contrast to the conventional EOS  $p=1/3(\rho-4B_{\rm eff})$ for non-interacting quark matter,  that taking the extreme strongly interacting limit on the unified IQM EOS gives $p=\rho-2B_{\rm eff}$, where  
$B_{\rm eff}$ is the effective bag constant. We employ the unified EOS to explore the properties of pure interacting quark stars (IQSs) composed of IQM. We describe how recent astrophysical observations, such as the pulsar-mass measurements, the NICER analysis, and the binary merger gravitational-wave events GW170817, GW190425, and GW190814, further constrain the parameter space. An upper bound for the maximum allowed mass of IQSs is found to be $M_{\rm TOV}\lesssim 3.23\, M_{\odot}$. Our analysis indicates a new possibility that the currently observed compact stars, including the recently reported GW190814's secondary component ($M=2.59^{+0.08}_{-0.09}\, M_{\odot}$), can be quark stars composed of interacting quark matter.
\end{abstract}
\maketitle

\section{Introduction}
It has been long expected that quark matter, a state consisting purely of quark and gluon degrees of freedom without confining into individual nucleons, can form at high densities or high temperatures~\cite{Ivanenko:1969gs,Itoh:1970uw,Collins:1974ky,Pasechnik:2016wkt,Pasechnik:2016wkt}. Bodmer~\cite{Bodmer:1971we}, Witten~\cite{Witten} and Terazawa~\cite{Terazawa:1979hq} proposed that quark matter with comparable numbers of $u, \,d, \,s$ quarks, also called strange quark matter (SQM), might be the ground state of baryonic matter at the zero temperature and pressure.  

However, a recent study~\cite{HRZ2017} demonstrated, taking the flavor-dependent feedback of the quark gas on the QCD vacuum into account, that $u, d$ quark matter ($ud$QM) is in general more stable than SQM, and can be more stable than the ordinary nuclei at sufficiently large baryon number beyond the periodic table. This has been connected to a series of recent phenomenological explorations~\cite{Zhang:2019mqb,Wang:2019jze,Wang:2019gam,Zhao:2019xqy,Iida:2020fyt,Xia:2020byy,Ren:2020tll,Miao:2020cqj,Cao:2020zxi} and experimental searches~\cite{Aad:2019pfm,Acharya:2020bed,Piotrowski:2020ftp}. 

Interacting quark matter (IQM) includes interquark effects from perturbative QCD (pQCD) and color superconductivity. pQCD corrections are due to the gluon-mediated interaction~\cite{Farhi:1984qu,Fraga:2001id,Fraga:2013qra}. Color superconductivity is the superconductivity in quark matter, arising from the spin-0 Cooper-pair condensation antisymmetric in color-flavor space~\cite{Alford:1998mk,Rajagopal:2000ff,Lugones:2002va}. This can result in two-flavor color superconductivity, where $u$ quarks pair with $d$ quarks 
[conventionally termed ``2SC" (``2SC+s") without (with)  strange quarks],\footnote{Another variant of two-flavor color superconductivity is the 2SCus phase, where $u$ pairs with $s$. Ref.~\cite{Alford:2002kj} showed that the 2SCus phase has the same free energy as the 2SC+s phase to order $m^4_s$, so we neglect the discussion of it in this paper.} or in a color-flavor locking (CFL) phase, where $u,d,s$ quarks pair with each other antisymmetrically. 

Most studies of color superconductivity have assumed an effective bag constant independent of the flavor composition,  resulting in the conclusion that 2SC phases are absent in compact star physics~\cite{Alford:2002kj}. However, improved Nambu–Jona-Lasinio models~\cite{Buballa:1998pr,Ratti:2002sh,OsipovSM} and quark-meson models~\cite{PWang01,HRZ2017} suggest that the effective bag constant  is very likely dependent on the flavor composition, opening up new possibilities.  A relatively small bag constant can make the two-flavor color superconductivity stable, warranting reconsideration of this possibility.

Binary mergers of compact stars produce gravitational waves, the waveforms of which encode  information about tidal deformation that is sensitive to the matter equation of state (EOS). In general, stars with stiff EOSs can easily be tidally deformed due to their large radii. The GW170817 event detected by LIGO/Virgo~\cite{TheLIGOScientific:2017qsa,Abbott:2018wiz} is the first confirmed merger event of compact stars. This, in conjunction with the more recent  GW190425 event~\cite{Abbott:2020uma}, has inspired many investigations into EOS and the gravitational properties of  nuclear matter~\cite{Abbott:2018exr, Radice:2017lry, Bauswein:2019skm,Annala:2017llu,De:2018uhw, Most:2018hfd, Most:2018eaw,Weih:2019rzo,Montana:2018bkb,Dexheimer:2019pay, Drago:2017bnf}, SQM~\cite{Zhou:2017pha,Burgio:2018yix}, and $ud$QM~\cite{Zhang:2019mqb,Ren:2020tll}.  


More recently, a binary merger event GW190814 was reported~\cite{Abbott:2020khf}, featuring a primary black hole with mass $23.2^{+1.1}_{+1.0} M_{\odot}$, and a secondary companion of $2.59^{+0.08}_{-0.09}\, M_{\odot}$, which is much larger than the upper bound   $M_{\rm TOV}\lesssim 2.3 M_{\odot}$ 
of the maximum mass of a non-rotating neutron star,
set by various analyses of GW170817~\cite{Margalit:2017dij,Rezzolla:2017aly,Ruiz:2017due,Shibata:2019ctb}. Conventional non-interacting SQM and $ud$QM 
have  bag constant values not sufficiently small to account for this large star mass~\cite{Zhou:2017pha, Zhang:2019mqb,Ren:2020tll}. It is consequently of interest to see how strongly interacting quark stars (IQSs), composed of IQM, can fit all these constraints.

We begin by first providing a unified framework for all  possible strongly interacting phases of SQM and $ud$QM by a simple reparametrization. After deriving a simple but universal EOS for IQM, we explore the properties of IQSs, investigating how recent astrophysical constraints such as   observed large pulsar masses~\cite{Demorest:2010bx,Antoniadis:2013pzd,   Cromartie:2019kug}, analysis of the NICER X-ray spectral-timing event data~\cite{Riley:2019yda,Miller:2019cac}, and the LIGO events~\cite{TheLIGOScientific:2017qsa,Abbott:2018wiz,Abbott:2020uma,Abbott:2020khf} constrain the IQS parameter space.  
\section{Properties of IQM}
We first rewrite the free energy $\Omega$ of the superconducting quark matter~\cite{Alford:2002kj} into a general form with the pQCD correction included~\cite{Fraga:2001id,Alford:2004pf,Weissenborn:2011qu}:  
\begin{equation}\begin{aligned}
\Omega=&-\frac{\xi_4}{4\pi^2}\mu^4+\frac{\xi_4(1-a_4)}{4\pi^2}\mu^4- \frac{ \xi_{2a} \Delta^2-\xi_{2b} m_s^2}{\pi^2}  \mu^2  \\
&-\frac{\mu_{e}^4}{12 \pi^2}+B_{\rm eff} ,
\label{omega_mu}
\end{aligned}\end{equation}
where $\mu$ and $\mu_e$ are the respective average quark and electron chemical potentials\footnote{For 2SC, $\mu=(\mu_u+2\mu_d)/3$. For 2SC+s and CFL phase, $\mu=(\mu_u+\mu_d+\mu_s)/3$.}. The first term represents the unpaired free quark gas contribution. The second term with $(1-a_4)$ represents the pQCD contribution from one-gluon exchange for gluon interaction to $O(\alpha_s^2)$ order. To phenomenologically account for  higher-order contributions, we can vary $a_4$ from $a_4=1$, corresponding to a vanishing pQCD correction, to very small values where these corrections become large~\cite{Fraga:2001id,Alford:2004pf,Weissenborn:2011qu}. The term with $m_s$ accounts for the correction from the finite strange quark mass if applicable, while the term with the gap parameter $\Delta$ represents the contribution from color superconductivity. The constant coefficients are
\begin{align}
(\xi_4,\xi_{2a}, \xi_{2b}) = \left\{ \begin{array} {ll}
(( \left(\frac{1}{3}\right)^{\frac{4}{3}}+ \left(\frac{2}{3}\right)^{\frac{4}{3}})^{-3},1,0) & \textrm{2SC phase}\\
(3,1,3/4) & \textrm{2SC+s phase}\\
(3,3,3/4)&   \textrm{CFL phase}
\end{array}
\nonumber
\right.
\end{align}
for the various types of quark matter.   $B_{\rm eff}$ is the effective bag constant that accounts for the non-perturbative contribution from the QCD vacuum, the size of which can be flavor-dependent~\cite{HRZ2017}. 

From the  thermodynamic relations
\be
p=-\Omega, \,\, n_{q}=-\frac{\partial\Omega}{\partial \mu},\,\, n_{e}=-\frac{\partial\Omega}{\partial \mu_e},\,\,   \rho=\Omega+n_q \mu+n_e \mu_e ,
\label{thermo}
\ee
 we obtain relevant thermodynamic quantities such as pressure $p$,  quark and electron number densities $n_{q,e}$, and   total energy density $\rho$, in terms of the chemical potential.  To reduce the size of the parameter space, we define
\be
\lambda=\frac{\xi_{2a} \Delta^2-\xi_{2b} m_s^2}{\sqrt{\xi_4 a_4}},
\label{lam}
\ee characterizing the strength of the related strong interaction.  The relations~(\ref{thermo}) then imply

\bea
n_q&=&\frac{\xi_4 a_4}{\pi^2}\mu^3 + \frac{\lambda
\sqrt{\xi_4 a_4} }{\pi^2}  2\mu, \quad  n_e=\frac{\mu_e^3}{3\pi^2},
\label{n_mu}
\eea
\bea
\rho&=&\frac{3\xi_4 a_4}{4\pi^2}\mu^4+\frac{\mu_{e}^4}{4 \pi^2}+B_{\rm eff} + \frac{ \lambda\sqrt{\xi_4 a_4} }{\pi^2}  \mu^2,
\label{rho_mu}
\eea
and using Eqs.~(\ref{omega_mu}) and (\ref{rho_mu}) we obtain 
\be
p=\frac{1}{3}(\rho-4B_{\rm eff})+ \frac{4\lambda^2}{9\pi^2}\left(-1+\text{sgn}{(\lambda)}\sqrt{1+3\pi^2 \frac{(\rho-B_{\rm eff})}{\lambda^2}}\right)
\label{eos_tot}
\ee
for the unified IQM EOS~\footnote{We removed the electron contribution in this derivation. A numerical check approves this approximation.}, where sgn$(\lambda)$ represents the sign of $\lambda$.
This general EOS expression above unifies the 2SC, ``2SC+s", and CFL phases. It only has 2 independent parameters $(B_{\rm eff},\lambda)$, while all  other parameters   $(a_4, \Delta,m_s)$ and $(\xi_4,\xi_{2a}, \xi_{2b})$ are subsumed in  $\lambda$ using Eq.~(\ref{lam})\footnote{As a necessary check of our general formula Eq.~(\ref{eos_tot}), inserting Eq.~(\ref{lam}) into it with the CFL factor where $(\xi_4,\xi_{2a}, \xi_{2b})=(3,3,3/4)$ can reproduce the CFL result in Ref.~\cite{Lugones:2002va, Pereira:2017rmp}.}.

We can further remove the $B_{\rm eff}$ parameter  by doing the dimensionless rescaling:
\be
\bar{\rho}=\frac{\rho}{4\,B_{\rm eff}}, \,\, \bar{p}=\frac{p}{4\,B_{\rm eff}},  \,\,
\label{scaling_prho}
\ee
and 
\be
 \bar{\lambda}=\frac{\lambda^2}{4B_{\rm eff}}= \frac{(\xi_{2a} \Delta^2-\xi_{2b} m_s^2)^2}{4\,B_{\rm eff}\xi_4 a_4},
 \label{scaling_lam}
\ee
so that the EOS Eq.~(\ref{eos_tot}) reduces to the dimensionless form 
\be
\bar{p}=\frac{1}{3}(\bar{\rho}-1)+ \frac{4}{9\pi^2}\bar{\lambda} \left(-1+\text{sgn}(\lambda)\sqrt{1+\frac{3\pi^2}{\bar{\lambda}} {(\bar{\rho}-\frac{1}{4})}}\right),
\label{eos_p}
\ee
In this paper, we only explore the positive $\lambda$ space.

With Eq.~(\ref{eos_p}), it is easy to show that $\partial \bar{p}/\partial \bar{\lambda}>0$, so a larger $\bar{\lambda}$ (i.e., smaller $B_{\rm eff}, a_4, m_s$ or larger $\Delta$) makes the EOS stiffer,  resulting in a larger star mass and radius. 

As $\bar{\lambda}\to0$, Eq.~(\ref{eos_p}) reduces to the conventional non-interacting rescaled quark matter EOS  $\bar{p}=(\bar{\rho}-1)/3$. At the opposite limit where $\bar{\lambda}$ is extremely large, Eq.~(\ref{eos_p}) approaches the special form
\be
\bar{p}\vert_{\bar{\lambda}\to \infty}=\bar{\rho}-\frac{1}{2}, 
\label{eos_infty}
\ee
which is equivalent to $p={\rho}-2B_{\rm eff}$ using Eq.~(\ref{scaling_prho}). We see that strong interaction effects can reduce the surface mass density of the quark star from $\rho_0= 4B_{\rm eff}$ down to $\rho_0=2B_{\rm eff}$, and increase the quark matter sound speed $c_s^2=\partial p/\partial \rho$ from $1/3$ up to $1$ (the light speed) maximally.

Since the energy per baryon number $E/A=3\mu\vert_{P=0}=3\mu\vert_{\Omega=0}$, we have
\be
\frac{E}{A}=\frac{3\sqrt{2} \pi}{(\xi_4 a_4)^{1/4}}\frac{ {B_{\rm eff}}^{1/4}}{\sqrt{\sqrt{4\bar{\lambda}+\pi^2}+2\sqrt{\bar{\lambda}}}}
\label{EA}
\ee
 from Eq.~(\ref{omega_mu}). 
We see that a smaller bag constant or a larger $\bar{\lambda}$ yields a smaller $E/A$.  As $\bar{\lambda} \to 0, \,a_4\to1$,   we recover the results for non-interacting quark matter~\cite{HRZ2017,Weber:2004kj}.
\section{Properties of IQS$s$}

\begin{figure}[h]
 \centering
\includegraphics[width=8cm]{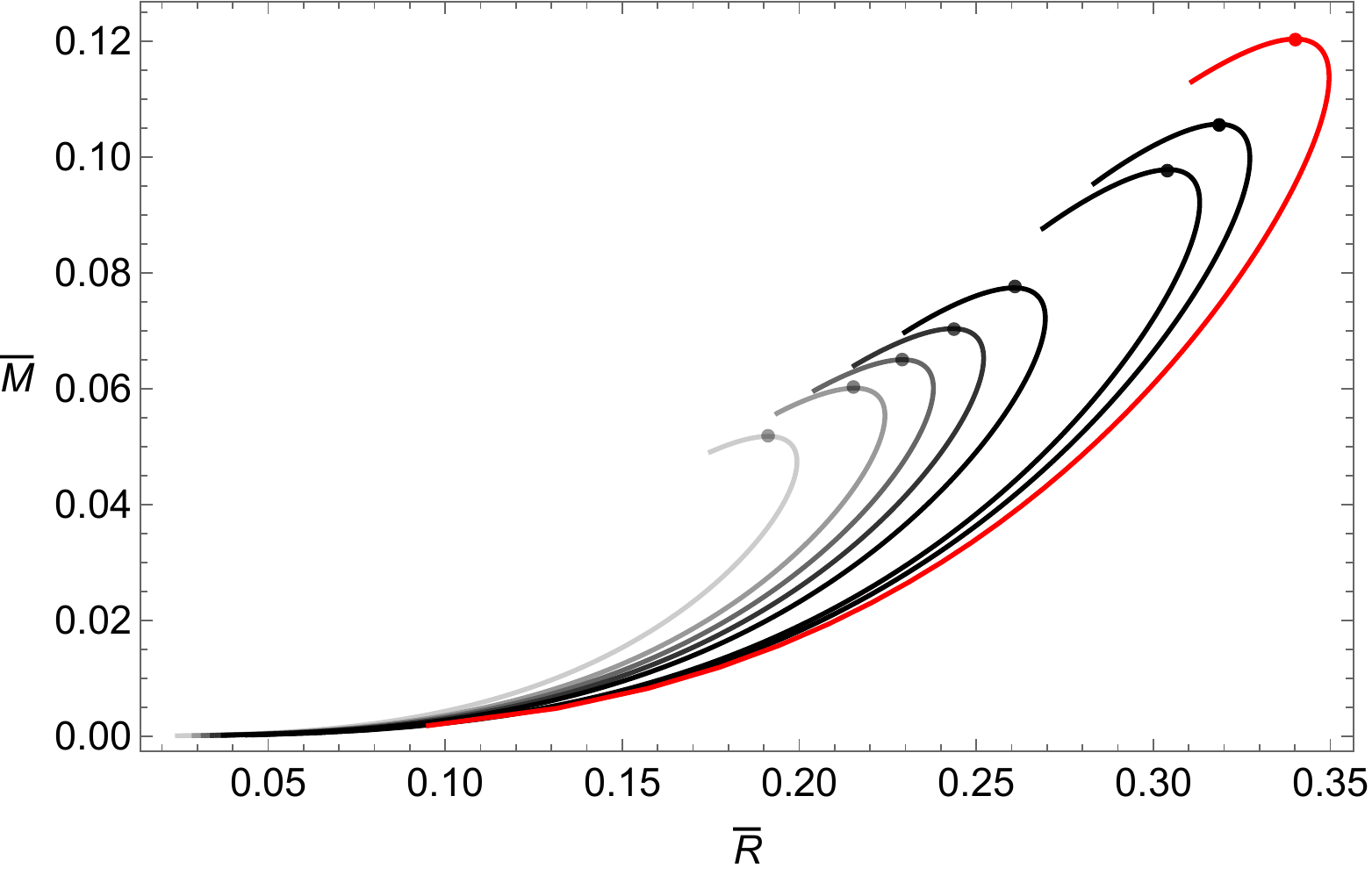}  
\caption{$\bar{M}$-$\bar{R}$ of strong-interacting quark stars for given $\bar{\lambda}$, sampling $(0,0.1, 0.25, 0.5, 1, 5, 10)$ from the lighter black line to the darker black line, respectively.  The red line corresponds to $\bar{\lambda}\to \infty$, with the corresponding EOS Eq.~(\ref{eos_infty}). The solid dots denote the maximum mass configurations for given $\bar{\lambda}$, for which $(\bar{M}_{\rm TOV}, \bar{R}_{\rm TOV})$ ranges from (0.05168, 0.1909) for $\bar{\lambda}=0$ to (0.1204,\,0.3400) for $\bar{\lambda}\to \infty$.}
   \label{rescaledMR}
\end{figure}

To study gravitational effects of interacting quark stars, we  further rescale the mass and radius into dimensionless form~\cite{Zdunik:2000xx,Haensel:2007yy}  in geometric units where $ c=G=1$
\be
 \bar{m}=m{\sqrt{4\,B_{\rm eff}}}, \quad \bar{r}={r}{\sqrt{4\,B_{\rm eff}}}, \,\,
\label{scaling_mr}
\ee
so that the Tolman-Oppenheimer-Volkov (TOV) equation~\cite{Oppenheimer:1939ne,Tolman:1939jz}
 \bea
 \begin{aligned}
{dp(r)\over dr}&=-{\left[m(r)+4\pi r^3p(r)\right]\left[\rho(r)+p(r)\right]\over r(r-2m(r))}\,,\,\,\\
{dm(r)\over dr}&=4\pi\rho(r)r^2,\, 
\end{aligned}
\label{tov}
\eea
can also be rescaled into dimensionless form (simply replace nonbarred symbols with barred ones). The rescaled TOV solution on $(\bar{M}, \bar{R}) =(M\sqrt{4B_{\rm eff}}, R\sqrt{4B_{\rm eff}})$ can thus be obtained with the rescaled EOS Eq.~(\ref{eos_p}) with respect to any given value of $\bar{\lambda}$.  
We depict the solutions to Eq.~\eqref{tov} in Fig.~\ref{rescaledMR} for the rescaled mass and radius for several different values of $\bar{\lambda}$. The physical $(M, R)$ with respect to any specific $B_{\rm eff}$ value can then straightforwardly be obtained directly from $(\bar{M}, \bar{R})$ using Eq.~(\ref{scaling_mr}).

From Fig.~\ref{rescaledMR}, we easily see that a larger $\bar{\lambda}$ leads to a larger $M_{\rm TOV}$ as expected, since a larger $\bar{\lambda}$ maps to a stiffer EOS. We can interpolate the $\bar{M}_{\rm TOV}(\bar{\lambda})$ numerical results with the following sigmoid-type function:
\be
\bar{M}_{\rm TOV}(\bar{\lambda})= \frac{\bar{M}_{\infty} }{1+c_1\, e^{-\bar{\lambda}^{c_2}}}+ \left(\bar{M}_0-  \frac{\bar{M}_{\infty}}{1+c_1\, e^{-\bar{\lambda}^{c_3}}} \right)e^{-\bar{\lambda}^{c_4}}
\label{Mbarmax}
\ee
where the coefficients $c_1\approx 0.8220 , c_2\approx0.4537, c_3\approx0.3313,c_4\approx0.2676$ are the best-fit values, with an error only at the $ 0.1\%$ level. And $\bar{M}_{0}=\bar{M}_{\rm TOV}(\bar{\lambda}\to 0)=0.05168$, $\bar{M}_{\infty}=\bar{M}_{\rm TOV}(\bar{\lambda}\to\infty)=0.1204$, corresponding to 
\bea
M_{\rm TOV}(\bar{\lambda}\to 0)&\approx&\frac{15.17\, M_{\odot}}{ (B_{\rm eff}/\rm MeV\, fm^{-3})^{1/2}},\\
M_{\rm TOV}(\bar{\lambda}\to \infty)&\approx&\frac{35.35\, M_{\odot}} {(B_{\rm eff}/\rm MeV\, fm^{-3})^{1/2}}
\eea
 referring to Eq.~(\ref{scaling_mr}). We see that the strongly interacting limit has a maximum star mass 2.33 times larger than that of the non-interacting case.
In general, we have the function $M_{\rm TOV} (B_{\rm eff}, \bar{\lambda})$ from Eq.~(\ref{Mbarmax}) after inserting Eq.~(\ref{scaling_mr}). The largest measured pulsar mass therefore imposes a constraint on the $(B_{\rm eff}, \bar{\lambda})$ space.

A recent NICER analysis of PSR J0030+0451~\cite{Riley:2019yda,Miller:2019cac} points to a star with a mass around $1.4\,M_{\odot}$ with a radius around 13 km ($90\%$ C.L.). The inferred contour of joint probability density distribution on the $M-R$ plane can then be translated to a range for constraints on the $(B_{\rm eff}, \bar{\lambda})$ space, utilizing the derived $\bar{M}(\bar{R})$ results presented in  Fig.~\ref{rescaledMR}. 

\section{Tidal deformability}
The response of compact stars to external disturbances is characterized by the Love number $k_2$~\cite{AELove,Hinderer:2007mb,Hinderer:2009ca,Postnikov:2010yn}:
\bea
\begin{aligned}
k_2 &= \frac{8 C^5}{5} (1-2C)^2 [ 2 + 2 C (y_R-1) -y_R ] \\
&\times \{ 2 C [ 6- 3y_R + 3C (5y_R-8)]+4C^3[ 13-11 y_R \\&+ C (3y_R-2)+2C^2(1+y_R)]  \\
&+ 3 (1-2C)^2 [ 2 - y_R + 2C (y_R-1)] \log (1-2C )\}^{-1} ,
\label{eqn:k2}
\end{aligned}
\eea 
where $C=M/R=C(\bar{M})$ for given $\bar{\lambda}$.  The quantity $y_R$ is $y(r)$ evaluated at the star surface, which can be obtained by solving the following equation \cite{Postnikov:2010yn}:
\bea
\begin{aligned}
&ry^\prime(r)+y(r)^2+ r^2Q(r) \\
&+y(r)e^{\lambda(r)}\left[1+4\pi r^2(p(r)-\rho(r))\right]=0\,,
\label{eqn:y}
\end{aligned}
\eea
with the boundary condition $y(0)=2$. Here
\bea
\begin{aligned}
Q(r)&=4\pi e^{\lambda(r)} \left(5\rho(r)+9p(r)+\frac{\rho(r)+p(r)}{c_s^2(r)} \right) \\
&-6\frac{e^{\lambda(r)}}{r^2}-\left(\nu^\prime(r)\right)^2,
\label{eq:Q}
\end{aligned}
\eea
\begin{equation}
e^{\lambda(r)}=\left[1-{2m(r)\over r}\right]^{-1} \quad
\nu^\prime(r)=2e^{\lambda(r)}{m(r)+4\pi p(r)r^3\over r^2}
\label{eq:met}
\end{equation}
and 
$c_s^2(r)\equiv dp/d\rho$ denotes the sound speed squared.
For stars with a finite surface density like quark stars, a matching condition~\cite{Damour:2009vw,Takatsy:2020bnx}  $y_R^{\rm ext}=y_R^{\rm int}-  4\pi R^3\rho_s/M=y_R^{\rm int}-  4\pi \bar{R}^3\bar{\rho}_s/\bar{M}$ should be used at the boundary. Solving Eq.~(\ref{eqn:y}) with  $\rho(r)$ and $p(r)$ obtained from Eq.~(\ref{tov}), we obtain the function $k_2(C)$ for a given $\bar{\lambda}$.  The dimensionless tidal deformability $\Lambda=2k_2/(3C^5)$ as a function of ($\bar{M}$, $\bar{\lambda}$) is thus obtained accordingly.  

\begin{figure}[h]
 \centering
\includegraphics[width=8cm]{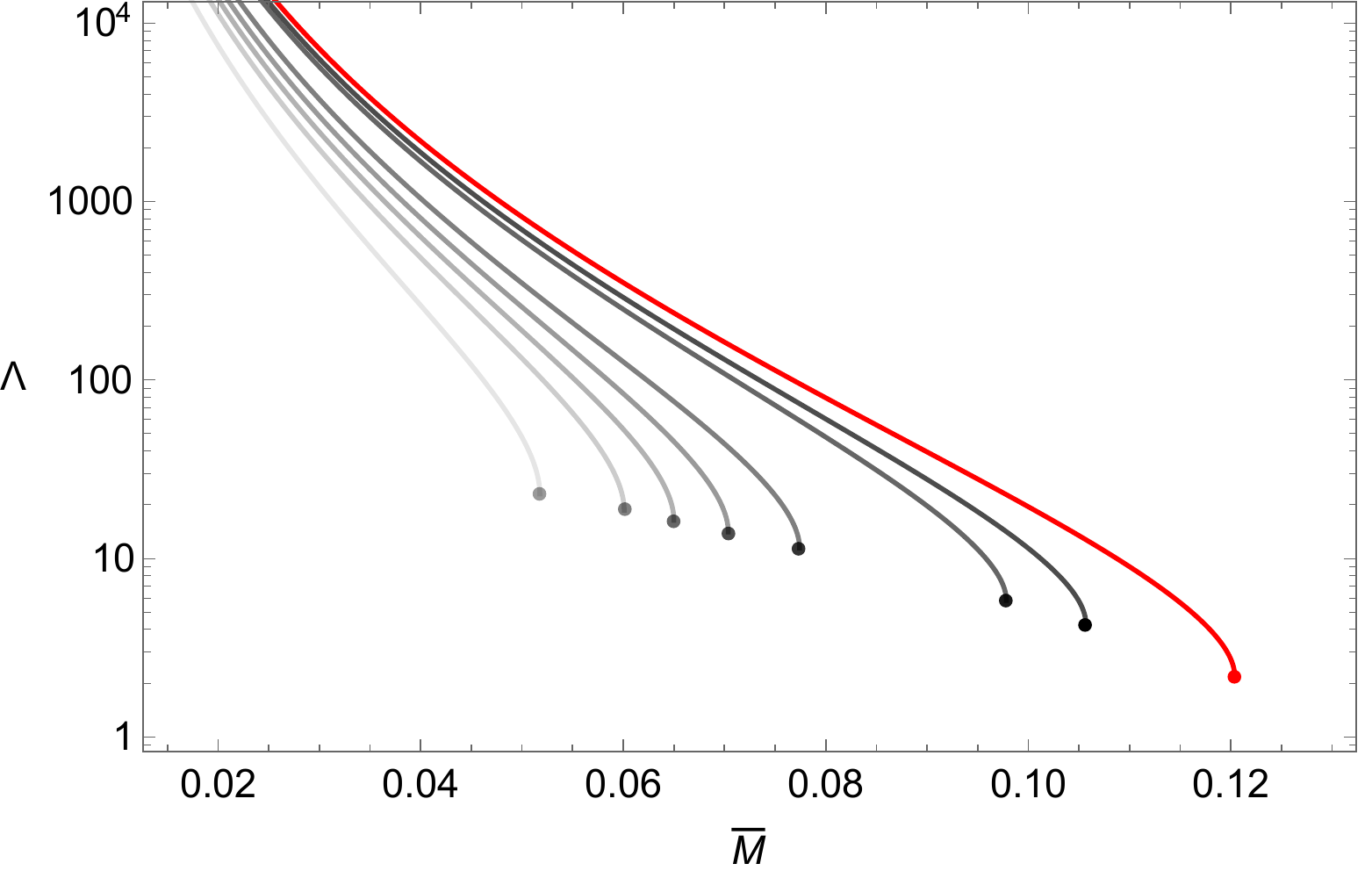}  
\caption{$\Lambda$-$\bar{M}$ of IQSs for $\bar{\lambda}=(0,0.1, 0.25, 0.5, 1, 5, 10)$, with a darker black color for a larger value. The red line corresponds to $\bar{\lambda}\to \infty$ utilizing the corresponding EOS~(\ref{eos_infty}). The solid dots denote the maximum mass configurations for the given $\bar{\lambda}$, with $(\bar{M}_{\rm TOV}, \Lambda_{\rm \bar{M}_{\rm TOV}})$ ranging from (0.0517, 22.9) for $\bar{\lambda}=0$ to (0.120,\,2.17) for $\bar{\lambda}\to \infty$.}
   \label{rescaledTidal}
\end{figure}

We depict the results of $\Lambda(\bar{M},\bar{\lambda})$  in Fig.~\ref{rescaledTidal}. Note that the lower end of each curve is determined by requiring the star mass not to exceed its maximum allowed mass. We can see that a larger $\bar{\lambda}$ results in a larger tidal deformability for a given $\bar{M}$, as expected since a larger $\bar{\lambda}$ maps to a stiffer EOS.

Assuming the compact objects detected by the recent LIGO event are pure IQSs, we can use the LIGO constraint on $\Lambda(1.4 M_\odot)$ to narrow down the parameter space $(B_{\rm eff},\bar{\lambda})$ with the $\Lambda(\bar{M},\bar{\lambda})$ calculated above. 

The average tidal deformability of a binary system is defined as
\bea
 \tilde \Lambda &=& \frac{16}{13} 
\frac{ (1+12q)}{(1+q)^5} {\Lambda} (M_1)+ \frac{16}{13}  \frac{q^4 (12+q)}{(1+q)^5}{\Lambda}({M}_2),
\label{LamLam}
\eea
where $M_1$ and $M_2$ are the masses of the binary components, and $q=M_2/M_1=\bar{M}_2/\bar{M}_1$, with $M_2$ being the smaller mass so that $0<q\lesssim1$. For any given chirp mass  $M_{c} = (M_1 M_2)^{3/5}/(M_1+M_2)^{1/5}$, one thus has
$
M_2=(q^2(q+1))^{1/5} M_c \text{ and }$ and $M_1=((1+q)/q^3)^{1/5} M_c\; .$
Using the rescaled $\bar{M}=M\sqrt{4 B_{\rm eff}}$, we eventually  obtain $\tilde \Lambda=\tilde \Lambda(\bar{M}_1,\bar{M}_2,\bar{\lambda})=\tilde \Lambda(M_c,q, B_{\rm eff},\bar{\lambda})$.  LIGO constraints on $\tilde{\Lambda}$ can then be used to narrow down the parameter space $(B_{\rm eff},\bar{\lambda})$ for given $M_c$ and $q$, assuming the objects detected by the recent LIGO events are pure IQSs. 

\begin{figure}[htb]  
 \centering
\includegraphics[width=8.5cm]{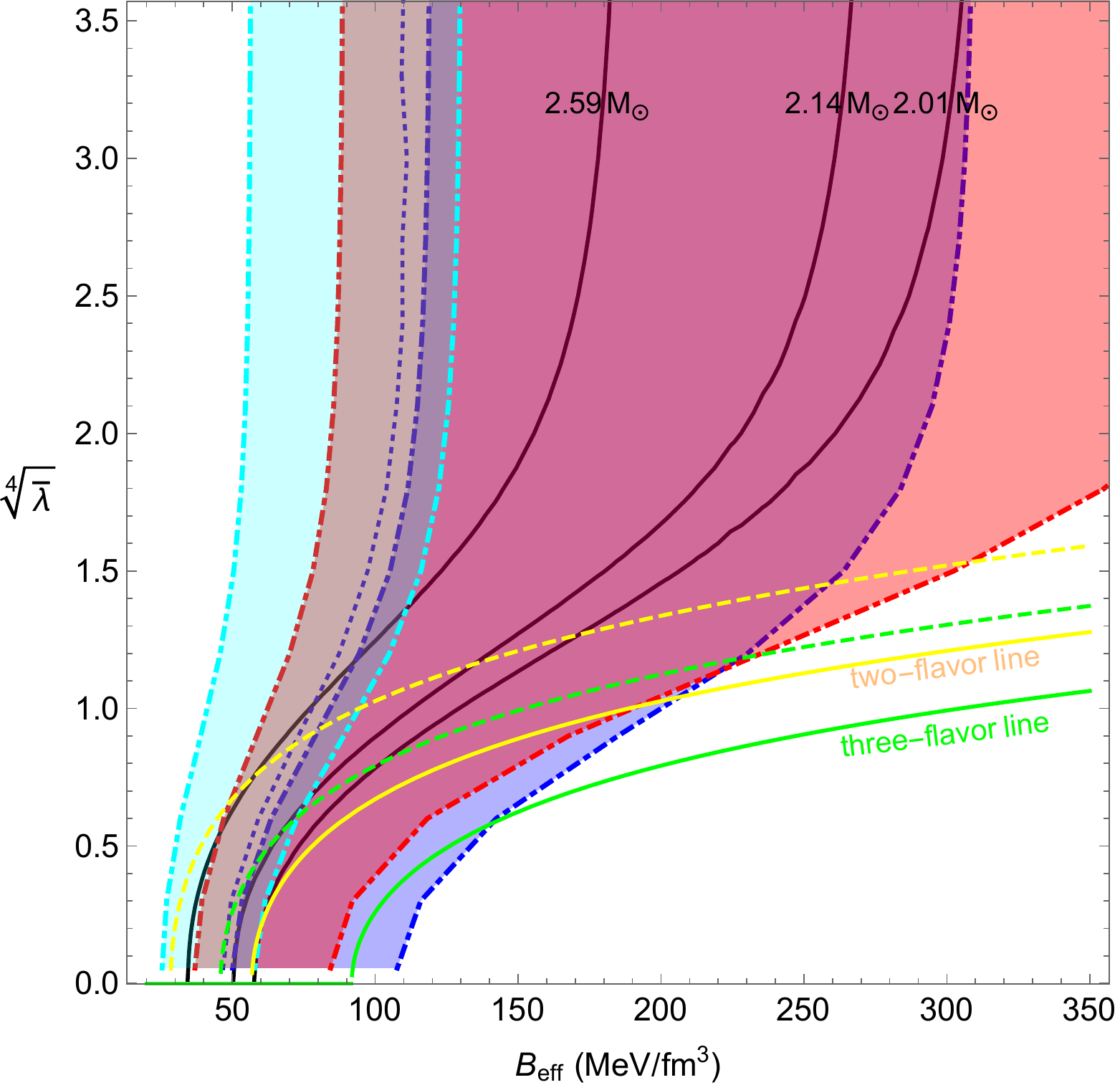}  
\caption{Astrophysical constraints on the parameter space in the $(B_{\rm eff}, \bar{\lambda}^{1/4})$ plane. 
Allowed regions are to the  left of each of the three solid black curves, which correspond to  maximum pulsar masses $M_{\rm TOV}\gtrsim 2.59 M_{\odot}$ (GW190814~\cite{Abbott:2020khf}), $2.14 M_{\odot}$(J0740+6620~\cite{ Cromartie:2019kug}), and $2.01 M_{\odot}$(J0348+0432~\cite{Antoniadis:2013pzd}), from left to right respectively. Colored dot-dash lines bound regions of the same color whose parameters satisfy various constraints ($90\%$ C.L.) : 
blue for  the GW170817 constraint $\tilde{\Lambda}=300^{+420}_{-230}$ with $M_c=1.186 \, M_{\odot}$  and $q=0.73-1.00$~\cite{TheLIGOScientific:2017qsa,Abbott:2018wiz}, red for GW190425 constraints $ \tilde{\Lambda} \lesssim 600$ with $M_c=1.44\, M_\odot$, $q=0.8-1.0$~\cite{Abbott:2020uma}, and   cyan for constraints from the recent NICER analysis of PSR J0030+0451~\cite{Miller:2019cac}; overlapping regions have correspondingly different colors.
The region to the right of the dotted blue line satisfies the GW170817 constraint $\Lambda_{1.4M_\odot}\lesssim 800$. 
 The yellow and green curves represent the stability lines derived from Eq.~(\ref{EA}), above which the two-flavor quark matter and three-flavor quark matter can be more stable than ordinary nuclei (i.e., $E/A <930\,\rm MeV$), respectively.  Dashed curves include the pQCD correction at $a_4=0.5$ order, whereas  the solid ones do not.}
\label{LamBarBcons}
\end{figure}

The resultant constrained parameter space is shown in Fig.~\ref{LamBarBcons}. Since we are studying quark stars, we do not use assumed hadronic EOS constraints. In our results, only the stability lines have an explicit dependence on the flavor composition and the size of $a_4$.

As described previously, the solid black curves are determined from Eq.~(\ref{Mbarmax}) for the measured pulsar masses.  The blue- and red-colored bands (bounded by dot-dashed lines) represent the $\tilde \Lambda$ constraint of GW170817 and GW190425 translated from the previously obtained $\tilde \Lambda(M_c,q, B_{\rm eff},\bar{\lambda})$ result. For a given $M_c$, the left edge of the band is determined by the upper bound of $\tilde{\Lambda}$ with the smallest allowed $q$ value, while the right edge is determined jointly by the lower bound of $\tilde{\Lambda}$ with $q=1$ and the requirement that $\bar{M}\lesssim\bar{M}_{\rm TOV}$.

Considering the solid black curve always goes leftward for a larger mass, there is then a critical mass value $M_c\approx3.23 M_{\odot}$ beyond which the associated solid black curve no longer intersects with the blue band (GW170817) for any $\bar{\lambda}$ value. Therefore, the possibility of IQSs is only allowed for $M_{\rm TOV}\lesssim 3.23\, M_{\odot}$.

One can easily observe that for any $\bar{\lambda}$ value, the lower bound of $B_{\rm eff}$ is determined by the LIGO constraint $\bar{\Lambda}\lesssim720$ of GW170817. The upper bound of $B_{\rm eff}$ is set by the constraint of $M_{\rm TOV}$ (the solid black curves)  for small $\bar{\lambda}$, and by the NICER constraint (the right edge of the cyan band) for large $\bar{\lambda}$. More explicitly, when $M_{\rm TOV}\gtrsim 2.59 M_{\odot}$ (the left of the left solid black curve) as suggested by GW190814,  the joint constraints tell that any  $B_{\rm eff}$ is excluded for $\bar{\lambda}^{1/4} \lesssim 1.17$, while $92.7 \lesssim B_{\rm eff}/ (\rm MeV/fm^3)\lesssim 111$ for $\bar{\lambda}^{1/4} \lesssim 1.37$, and $111 \lesssim B_{\rm eff}/ (\rm MeV/fm^3)\lesssim 130$ for $\bar{\lambda}^{1/4} \gtrsim 1.37$.  For comparison, relaxing the maximum mass to $M_{\rm TOV}\gtrsim 2.14\, M_{\odot}$ (the left of the middle solid black curve), we have the constrained parameter space $49.5 \lesssim B_{\rm eff}/ (\rm MeV/fm^3) \lesssim 76.2$ for  $\bar{\lambda}^{1/4} \lesssim 0.67$, and $76.2 \lesssim B_{\rm eff}/ (\rm MeV/fm^3)\lesssim 130$ for $\bar{\lambda}^{1/4} \gtrsim 0.67$. 
We see that the constrained region can be well above the two-flavor and three-flavor stability lines within reasonable uncertainties of $a_4$, so that both interacting $ud$QM and interacting SQM can be more stable than ordinary nuclei in the allowed parameter space. 

\section{Summary}

We have explored IQM, i.e., quark matter with interquark effects from pQCD corrections and color superconductivity, in a general parametrization, reducing the EOS to a dimensionless form depending on one single parameter $\bar{\lambda}$, which characterizes the relative size of strong interaction effects. A larger $\bar{\lambda}$ results in a stiffer EOS. At the large $\bar{\lambda}$ limit, the EOS becomes $\bar{p}=\bar{\rho}-1/2$, or equivalently $p=\rho-2B_{\rm eff}.$ 


Using this we studied the properties of IQSs (compact stars composed of IQM) and found that a larger $\bar{\lambda}$ results in a larger (rescaled) maximum mass, with an upper bound $M_{\rm TOV}(\bar{\lambda}\to \infty)=35.35\, M_{\odot} (B_{\rm eff}/\rm MeV\, fm^{-3})^{-1/2}$. We also obtained an explicit expression of $\bar{M}_{\rm TOV}(\bar{\lambda})$, and computed the deformability of IQSs. 

Assuming the related compact objects are IQSs, we translated recent astrophysical observations into various constraints on the  $(B_{\rm eff}, \bar{\lambda}^{1/4})$ parameter space, and obtained an upper bound for the maximum allowed mass for IQSs: $M_{\rm TOV}\lesssim 3.23 M_{\odot}$. Remarkably, the parameter space is confined to a window of moderate $B_{\rm eff}$ and large $\bar{\lambda}$ above the two-flavor stability line, enclosed by the constraints from the upper bound of $\tilde \Lambda$ of GW170817, the recent NICER analysis, and the mass of the most massive compact star identified. Compact stars identified using the recent astrophysical observations can thus be consistently interpreted as quark stars composed of interacting $ud$QM or  interacting SQM. This study paves the way for future astrophysical observations to confirm this possibility, further constrain it,  or rule it out entirely.

\begin{acknowledgments}
\noindent\textbf{Acknowledgments. }  
We thank Bob Holdom and Jing Ren for helpful discussions. This research is supported in part by the Natural Sciences and Engineering Research Council of Canada. 
\end{acknowledgments}


\begin{thebibliography}{}
%
\bibitem{Ivanenko:1969gs} 
D.~Ivanenko and D.~F.~Kurdgelaidze,
Lett. Nuovo Cim. \textbf{2}, 13-16 (1969)

\bibitem{Itoh:1970uw}
N.~Itoh,
Prog. Theor. Phys. \textbf{44}, 291 (1970)

\bibitem{Collins:1974ky}
J.~C.~Collins and M.~J.~Perry,
Phys. Rev. Lett. \textbf{34}, 1353 (1975)
\bibitem{Pasechnik:2016wkt}
R.~Pasechnik and M.~\v{S}umbera,
Universe \textbf{3}, 7 (2017)
[arXiv:1611.01533 [hep-ph]].

\bibitem{Bodmer:1971we} 
  A.~R.~Bodmer,
  Phys.\ Rev.\ D {\bf 4}, 1601 (1971).
\bibitem{Witten}
  E.~Witten,
  Phys.\ Rev.\ D {\bf 30}, 272 (1984).
\bibitem{Terazawa:1979hq} 
  H.~Terazawa,
  INS-Report-336 (INS, University of Tokyo, Tokyo) May, 1979.

 \bibitem{HRZ2017}
B.~Holdom, J.~Ren and C.~Zhang,
Phys. Rev. Lett. \textbf{120}, 222001 (2018)
[arXiv:1707.06610 [hep-ph]].


\bibitem{Zhang:2019mqb}
C.~Zhang,
Phys. Rev. D \textbf{101}, 043003 (2020)
[arXiv:1908.10355 [astro-ph.HE]].
\bibitem{Wang:2019jze}
Q.~Wang, C.~Shi and H.~S.~Zong,
Phys. Rev. D \textbf{100}, 123003 (2019)
[arXiv:1908.06558 [hep-ph]].

\bibitem{Wang:2019gam}
Q.~Wang, T.~Zhao and H.~Zong,
[arXiv:1908.01325 [hep-ph]].
\bibitem{Zhao:2019xqy}
T.~Zhao, W.~Zheng, F.~Wang, C.~M.~Li, Y.~Yan, Y.~F.~Huang and H.~S.~Zong,
Phys. Rev. D \textbf{100}, 043018 (2019)
[arXiv:1904.09744 [nucl-th]].
\bibitem{Xia:2020byy}
C.~J.~Xia, S.~S.~Xue, R.~X.~Xu and S.~G.~Zhou,
Phys. Rev. D \textbf{101}, 103031 (2020)
[arXiv:2001.03531 [nucl-th]].
\bibitem{Iida:2020fyt}
K.~Iida and T.~Fujie,
JPS Conf. Proc. \textbf{31}, 011057 (2020)


\bibitem{Ren:2020tll}
J.~Ren and C.~Zhang,
[arXiv:2006.09604 [hep-ph]].

\bibitem{Miao:2020cqj}
Z.~Q.~Miao, C.~J.~Xia, X.~Y.~Lai, T.~Maruyama, R.~X.~Xu and E.~P.~Zhou,
[arXiv:2008.06932 [nucl-th]].

\bibitem{Cao:2020zxi}
Z.~Cao, L.~W.~Chen, P.~C.~Chu and Y.~Zhou,
[arXiv:2009.00942 [astro-ph.HE]].

 
\bibitem{Aad:2019pfm} 
  G.~Aad {\it et al.} (ATLAS Collaboration),
  arXiv:1905.10130 [hep-ex].
  
\bibitem{Acharya:2020bed}
B.~Acharya \textit{et al.} (MoEDAL Collaboration),
Phys. Rev. Lett. \textbf{126}, 071801 (2021)
doi:10.1103/PhysRevLett.126.071801
[arXiv:2002.00861 [hep-ex]].

\bibitem{Piotrowski:2020ftp}
L.~W.~Piotrowski, K.~Malek, L.~Mankiewicz, M.~Sokolowski, G.~Wrochna, A.~Zadrozny and A.~F.~Zarnecki,
Phys. Rev. Lett. \textbf{125}, 091101 (2020)
[arXiv:2008.01285 [astro-ph.HE]].

\bibitem{Farhi:1984qu}
E.~Farhi and R.~L.~Jaffe,
Phys. Rev. D \textbf{30}, 2379 (1984)
    
\bibitem{Fraga:2001id}
E.~S.~Fraga, R.~D.~Pisarski and J.~Schaffner-Bielich,
Phys. Rev. D \textbf{63}, 121702 (2001)
[arXiv:hep-ph/0101143 [hep-ph]].
\bibitem{Fraga:2013qra}
E.~S.~Fraga, A.~Kurkela and A.~Vuorinen,
Astrophys. J. Lett. \textbf{781}, L25 (2014)
[arXiv:1311.5154 [nucl-th]].
  
\bibitem{Alford:1998mk}
M.~G.~Alford, K.~Rajagopal and F.~Wilczek,
Nucl. Phys. B \textbf{537}, 443-458 (1999)
[arXiv:hep-ph/9804403 [hep-ph]].
  
\bibitem{Rajagopal:2000ff}
K.~Rajagopal and F.~Wilczek,
Phys. Rev. Lett. \textbf{86}, 3492-3495 (2001)
[arXiv:hep-ph/0012039 [hep-ph]].
\bibitem{Lugones:2002va}
G.~Lugones and J.~E.~Horvath,
Phys. Rev. D \textbf{66}, 074017 (2002)
[arXiv:hep-ph/0211070 [hep-ph]].
   
     \bibitem{Alford:2002kj} 
  M.~Alford and K.~Rajagopal,
  JHEP {\bf 0206}, 031 (2002)
  [hep-ph/0204001].

\bibitem{Buballa:1998pr} 
  M.~Buballa and M.~Oertel,
  Phys.\ Lett.\ B {\bf 457}, 261 (1999)
  [hep-ph/9810529].

\bibitem{Ratti:2002sh} 
  C.~Ratti,
  Europhys.\ Lett.\  {\bf 61}, 314 (2003)
  [hep-ph/0210295].
\bibitem{OsipovSM}
J.~Moreira, J.~Morais, B.~Hiller, A.~A.~Osipov and A.~H.~Blin,
  Phys.\ Rev.\ D {\bf 91}, 116003 (2015)
  [arXiv:1409.0336 [hep-ph]].

\bibitem{PWang01}
  P.~Wang, V.~E.~Lyubovitskij, T.~Gutsche and A.~Faessler,
  Phys.\ Rev.\ C {\bf 67} (2003) 015210
  [hep-ph/0205251].


\bibitem{TheLIGOScientific:2017qsa}
B.~Abbott \textit{et al.} (LIGO Scientific and Virgo Collaborations),
Phys. Rev. Lett. \textbf{119}, 161101 (2017)
[arXiv:1710.05832 [gr-qc]].
  \bibitem{Abbott:2018wiz} 
  B.~P.~Abbott {\it et al.} [LIGO Scientific and Virgo Collaborations],
  Phys.\ Rev.\ X {\bf 9}, 011001 (2019)
  [arXiv:1805.11579 [gr-qc]].
  
\bibitem{Abbott:2020uma}
B.~Abbott \textit{et al.} (LIGO Scientific and Virgo Collaborations),
Astrophys. J. Lett. \textbf{892}, L3 (2020)
[arXiv:2001.01761 [astro-ph.HE]].



\bibitem{Abbott:2018exr} 
  B.~P.~Abbott {\it et al.} (LIGO Scientific and Virgo Collaborations),
  Phys.\ Rev.\ Lett.\  {\bf 121}, 161101 (2018)
  [arXiv:1805.11581 [gr-qc]].
  
   \bibitem{Radice:2017lry} 
  D.~Radice, A.~Perego, F.~Zappa and S.~Bernuzzi,
  Astrophys.\ J.\  {\bf 852}, L29 (2018)
  [arXiv:1711.03647 [astro-ph.HE]].
  \bibitem{Bauswein:2019skm} 
  A.~Bauswein {\it et al.},
  AIP Conf.\ Proc.\  {\bf 2127},  020013 (2019)
  [arXiv:1904.01306 [astro-ph.HE]].
 
\bibitem{Annala:2017llu} 
  E.~Annala, T.~Gorda, A.~Kurkela and A.~Vuorinen,
  Phys.\ Rev.\ Lett.\  {\bf 120}, 172703 (2018)
  [arXiv:1711.02644 [astro-ph.HE]].
  
  
\bibitem{De:2018uhw} 
  S.~De, D.~Finstad, J.~M.~Lattimer, D.~A.~Brown, E.~Berger and C.~M.~Biwer,
  Phys.\ Rev.\ Lett.\  {\bf 121}, 091102 (2018)
  Erratum: [Phys.\ Rev.\ Lett.\  {\bf 121}, 259902 (2018)]
  [arXiv:1804.08583 [astro-ph.HE]].
  
 \bibitem{Drago:2017bnf} 
  A.~Drago and G.~Pagliara,
  Astrophys.\ J.\  {\bf 852}, L32 (2018)
  [arXiv:1710.02003 [astro-ph.HE]].
  

 
\bibitem{Most:2018hfd} 
  E.~R.~Most, L.~R.~Weih, L.~Rezzolla and J.~Schaffner-Bielich,
  Phys.\ Rev.\ Lett.\  {\bf 120},  261103 (2018)
  [arXiv:1803.00549 [gr-qc]].

\bibitem{Most:2018eaw} 
  E.~R.~Most, L.~J.~Papenfort, V.~Dexheimer, M.~Hanauske, S.~Schramm, H.~St\"ocker, and L.~Rezzolla,
  Phys.\ Rev.\ Lett.\  {\bf 122}, 061101 (2019)
  [arXiv:1807.03684 [astro-ph.HE]].

\bibitem{Weih:2019rzo} 
  L.~R.~Weih, E.~R.~Most and L.~Rezzolla,
  Astrophys.\ J.\  {\bf 881}, 73 (2019)
  [arXiv:1905.04900 [astro-ph.HE]].
  
\bibitem{Montana:2018bkb} 
  G.~Montana, L.~Tolos, M.~Hanauske and L.~Rezzolla,
  Phys.\ Rev.\ D {\bf 99}, 103009 (2019)
  [arXiv:1811.10929 [astro-ph.HE]].

\bibitem{Dexheimer:2019pay} 
  V.~Dexheimer, L.~T.~T.~Soethe, J.~Roark, R.~O.~Gomes, S.~O.~Kepler and S.~Schramm,
  Int.\ J.\ Mod.\ Phys.\ E {\bf 27}, 1830008 (2018)
  [arXiv:1901.03252 [astro-ph.HE]].

  \bibitem{Zhou:2017pha} 
  E.~P.~Zhou, X.~Zhou and A.~Li,
  Phys.\ Rev.\ D {\bf 97}, 083015 (2018)
  [arXiv:1711.04312 [astro-ph.HE]].
  
  
    \bibitem{Burgio:2018yix} 
  G.~F.~Burgio, A.~Drago, G.~Pagliara, H.~J.~Schulze and J.~B.~Wei,
  Astrophys.\ J.\  {\bf 860}, 139 (2018)
  [arXiv:1803.09696 [astro-ph.HE]].

\bibitem{Abbott:2020khf}
R.~Abbott \textit{et al.} [LIGO Scientific and Virgo Collaborations],
Astrophys. J. Lett. \textbf{896}, L44 (2020)
[arXiv:2006.12611 [astro-ph.HE]].

\bibitem{Margalit:2017dij}
B.~Margalit and B.~D.~Metzger,
Astrophys. J. Lett. \textbf{850}, L19 (2017)
[arXiv:1710.05938 [astro-ph.HE]].

\bibitem{Rezzolla:2017aly}
L.~Rezzolla, E.~R.~Most and L.~R.~Weih,
Astrophys. J. Lett. \textbf{852}, L25 (2018)
[arXiv:1711.00314 [astro-ph.HE]].

\bibitem{Ruiz:2017due}
M.~Ruiz, S.~L.~Shapiro and A.~Tsokaros,
Phys. Rev. D \textbf{97}, 021501 (2018)
[arXiv:1711.00473 [astro-ph.HE]].

\bibitem{Shibata:2019ctb}
M.~Shibata, E.~Zhou, K.~Kiuchi and S.~Fujibayashi,
Phys. Rev. D \textbf{100}, 023015 (2019)
[arXiv:1905.03656 [astro-ph.HE]].





   \bibitem{Demorest:2010bx} 
  P.~Demorest, T.~Pennucci, S.~Ransom, M.~Roberts and J.~Hessels,
  Nature {\bf 467}, 1081 (2010)
  [arXiv:1010.5788 [astro-ph.HE]].


\bibitem{Antoniadis:2013pzd}
J.~Antoniadis  \textit{et al.},
Science \textbf{340}, 1233232 (2013)
[arXiv:1304.6875 [astro-ph.HE]].


\bibitem{Cromartie:2019kug}
H.~T.~Cromartie \textit{et al.} [NANOGrav],
Nature Astron. \textbf{4}, 72 (2020)
[arXiv:1904.06759 [astro-ph.HE]].
   
   \bibitem{Riley:2019yda}
T.~E.~Riley  \textit{et al.},
Astrophys. J. Lett. \textbf{887}, L21 (2019)
[arXiv:1912.05702 [astro-ph.HE]].
   \bibitem{Miller:2019cac}
M.~Miller  \textit{et al.},
Astrophys. J. Lett. \textbf{887}, L24 (2019)
[arXiv:1912.05705 [astro-ph.HE]].



\bibitem{Alford:2004pf}
M.~Alford, M.~Braby, M.~W.~Paris and S.~Reddy,
Astrophys. J. \textbf{629}, 969-978 (2005)
[arXiv:nucl-th/0411016 [nucl-th]].
\bibitem{Weissenborn:2011qu}
S.~Weissenborn, I.~Sagert, G.~Pagliara, M.~Hempel and J.~Schaffner-Bielich,
Astrophys. J. Lett. \textbf{740}, L14 (2011)
[arXiv:1102.2869 [astro-ph.HE]].

\bibitem{Pereira:2017rmp}
J.~P.~Pereira, C.~V.~Flores and G.~Lugones,
Astrophys. J. \textbf{860}, 12 (2018)
[arXiv:1706.09371 [gr-qc]].

\bibitem{Weber:2004kj}
F.~Weber,
Prog. Part. Nucl. Phys. \textbf{54}, 193-288 (2005)
[arXiv:astro-ph/0407155 [astro-ph]].

\bibitem{Zdunik:2000xx} 
  J.~L.~Zdunik,
  Astron.\ Astrophys.\  {\bf 359}, 311 (2000)
  [astro-ph/0004375].
\bibitem{Haensel:2007yy} 
  P.~Haensel, A.~Y.~Potekhin and D.~G.~Yakovlev,
  Astrophys.\ Space Sci.\ Libr.\  {\bf 326}, pp.1 (2007).
  
\bibitem{Tolman:1939jz} 
  R.~C.~Tolman,
  Phys.\ Rev.\  {\bf 55}, 364 (1939).
  
\bibitem{Oppenheimer:1939ne} 
  J.~R.~Oppenheimer and G.~M.~Volkoff,
  Phys.\ Rev.\  {\bf 55}, 374 (1939).


    \bibitem{AELove} A. E. H. Love, Proc. R. Soc. A 82, 73 (1909).
    \bibitem{Hinderer:2007mb} 
  T.~Hinderer,
  Astrophys.\ J.\  {\bf 677}, 1216 (2008)
  [arXiv:0711.2420 [astro-ph]].
\bibitem{Hinderer:2009ca} 
  T.~Hinderer, B.~D.~Lackey, R.~N.~Lang and J.~S.~Read,
  Phys.\ Rev.\ D {\bf 81}, 123016 (2010)
  [arXiv:0911.3535 [astro-ph.HE]].

  \bibitem{Postnikov:2010yn} 
  S.~Postnikov, M.~Prakash and J.~M.~Lattimer,
  Phys.\ Rev.\ D {\bf 82}, 024016 (2010)
  [arXiv:1004.5098 [astro-ph.SR]].
  
  \bibitem{Damour:2009vw} 
  T.~Damour and A.~Nagar,
  Phys.\ Rev.\ D {\bf 80}, 084035 (2009)
  [arXiv:0906.0096 [gr-qc]].

\bibitem{Takatsy:2020bnx}
J.~Takatsy and P.~Kovacs,
Phys. Rev. D \textbf{102}, 028501 (2020)
[arXiv:2007.01139 [astro-ph.HE]].

  
\end{thebibliography}
\end{document}